\documentclass[twocolumn,showpacs,preprintnumbers,a4paper,fleqn,pre]{revtex4}
\usepackage{amssymb}
\usepackage{amsfonts}
\usepackage{amsmath}
\usepackage{graphicx}
\usepackage{dcolumn}
\usepackage{bm}

\input{tcilatex}

\begin{document}

\preprint{APS/123-QED}
\title{Surrogate Test to Distinguish between Chaotic and Pseudoperiodic Time
Series}
\author{Xiaodong Luo }
\email{enxdluo@eie.polyu.edu.hk}
\author{Tomomichi Nakamura}
\author{Michael Small}
\affiliation{Department of Electronic and Information Engineering, Hong Kong Polytechnic
University, Hung Hom, Hong Kong.}
\date{\today }

\begin{abstract}
In this communication a new algorithm is proposed to produce surrogates for
pseudoperiodic time series. By imposing a few constraints on the noise
components of pseudoperiodic data sets, we devise an effective method to
generate surrogates. Unlike other algorithms, this method properly copes
with pseudoperiodic orbits contaminated with linear colored observational
noise. We will demonstrate the ability of this algorithm to distinguish
chaotic orbits from pseudoperiodic orbits through simulation data sets from
the 
R\"{o}ssler
system. As an example of application of this algorithm, we will also employ
it to investigate a human electrocardiogram (ECG) record.
\end{abstract}

\maketitle

\section{Introduction}

Surrogate tests \cite{Theiler testing} are examples of Monte Carlo
hypothesis tests \cite{Galka topics}. Taking the surrogate test of
nonlinearity in a time series \cite{Theiler testing} as an example, we first
need to adopt a null hypothesis, which usually supposes the time series is
generated by a linear stochastic process and potentially filtered by a
nonlinear filter \cite{note nonlinearity}. Based on this null hypothesis, a
large number of data sets (surrogates) are to be produced from the original
time series, which keeps the linearity of the original time series but
destroys all other structures. We then calculate some nonlinear statistics
(discriminating statistics), for example, correlation dimension, of both the
original time series and the surrogates. If the discriminating statistic of
the original time series deviates from those of the surrogates, we can
reject the null hypothesis we proposed and claim that the original time
series is deterministic with certain confidence level (depending on how many
surrogates we have generated, to be shown later). In general, to apply the
surrogate technique to test if a time series possesses the property $P$ we
are interested, we first select a null hypothesis, which assumes the time
series instead has a property $Q$ opposite to $P$. We then devise a
corresponding algorithm to produce surrogates from the observed data set. In
principle, these surrogates shall preserve the potential property $Q$ while
destroying all others. The next step is to choose a suitable discriminating
statistic, which shall be an invariant measure for both the surrogates and
the original time series if the null hypothesis is true. Hence if the
discriminating statistic of the original time series distinctly deviates
from the distribution of the discriminating statistic of the surrogates, the
null hypothesis is unlikely to be true, or in other words, the time series
is much more likely to possess the property $P$ than $Q$. In this way, we
can assess the statistical significance of our calculations through
surrogate test technique even when we have only a very limited amount of
observations. Such assessments are important because in many practical
situations statistical fluctuations are inevitable due to the presence of
noise, hence the surrogate test is a proper tool to evaluate the reliability
of our results in a statistical sense.

In this communication, we are focused on discussing the algorithm to
generate surrogates for pseudoperiodic time series. By pseudoperiodic time
series we mean a representative of a periodic orbit perturbed by dynamical
noise, or contaminated by observational noise, or with the combination of
the both noises, whose states within one cycle are largely independent of
those within previous cycles given a cycle length. Note that, in our
discussions we will always assume we have detected that the time series are
produced from nonlinear deterministic systems, but they are also possibly
contaminated by some noises. As we know, if an irregular time series comes
from a nonlinear deterministic system, it shall be either chaotic or
pseudoperiodic in most cases. In some situations, it might be important for
us to discriminate between pseudoperiodicity and chaos. However, chaotic and
pseudoperiodic time series often look similar, we might not be able to
distinguish them from each other only through visual inspections,
quantitative techniques are needed instead at this time. One choice is to
apply the direct test techniques. For instance, we can calculate some
characteristic statistics of the time series, such as the Lyapunov exponent
and the correlation dimension. However, a direct test usually will not give
out the confidence level. If we find the Lyapunov exponent of a time series
is, for example, $0.01$, it may be difficult for us to tell whether the time
series is chaotic or the time series is pseudoperiodic, but the presence of
noise causes the Lyapunov exponent to be slightly larger than zero. As an
alternative choice, we suggest one utilizes the surrogate test rather than
the direct test, which can provide us the confidence level by calculating a
large number of surrogates. Through the surrogate tests, if we could exclude
the possibility that the time series is pseudoperiodic, then the time series
is more likely to be chaotic. This is the essential idea to apply our
algorithm to distinguish chaos from pseudoperiodicity, as to be shown in
section III.

First let us briefly review some of the algorithms to generate surrogates
for pseudoperiodic time series. Initially, to generate surrogates for
pseudoperiodic time series, Theiler \cite{Theiler on} proposed the cycle
shuffling algorithm. The idea is to divide the whole data set into some
segments and let each segment contain exactly an integer number of cycles.
The surrogates are obtained by randomly shuffling these segments, which will
preserve the intracycle dynamics but destroy the intercycle ones by
randomizing the temporal sequence of the individual cycles. The difficulty
in applying this algorithm is that it requires preknowledge of the precise
periodicity, otherwise shuffling the individual cycles might lead to
spurious results \cite{Small surrogate}.

Recently, with the development of the cyclic theory of chaos \cite{Ayerbach
exploring}, many authors have shown interest in searching unstable periodic
orbits (UPOs) in noisy data sets from chaotic dynamical systems. The
algorithms proposed in \cite{Pierson detecting} essentially deal with the
unstable fixed points of the UPOs. But as observed, the presence of noise
will reduce the statistical significance of these algorithms. One remedy is
to introduce the surrogate test for reliability assessments, e.g., Dolan 
\textit{et.al} \cite{Pierson detecting} claimed that the randomly shuffling
surrogate algorithm \cite{Theiler testing} together with the simple
recurrence method \cite{Pierson detecting} correctly tests the appropriate
null hypothesis. Essentially, this approach is very similar to the cycle
shuffling algorithm described previously. The simple recurrence algorithm is
equivalent to applying a Poincar\'{e} map on the continuous dynamical
systems and then studying only the data points falling on the cross-section
plane, hence one does not need to consider the intracycle dynamics and no
knowledge of the periodicity is required, while randomly shuffling these
data points exactly aims to randomize the temporal sequence of the cycles.
However, one potential problem of this algorithm is that it might generate
spuriously high statistical significance due to the correlation between the
cycles \cite{Petracchi the}.

Later, Small \textit{et.al }\cite{Small surrogate}\textit{\ }proposed the
pseudoperiodic surrogate (PPS) algorithm from another viewpoint. They first
apply the time delay embedding reconstruction \cite{Takens detecting} to the
original data set, then utilize a method based on local linear modelling
techniques to produce surrogate data which approximate the behavior of the
underlying dynamical system. As the authors pointed out, this algorithm
works well even with very large dynamical noise, but it may incorrectly
reject the null hypothesis if the intercycles of the pseudoperiodic orbit
have a linear stochastic dependence induced by colored additive
observational noise \cite{note noise}.

In this communication we propose a new surrogate algorithm for continuous
dynamical systems, which properly copes with linear stochastic dependence
between the cycles of the pseudoperiodic orbits. The null hypothesis to be
tested is that the stationary data set is pseudoperiodic with noise
components which are (approximately) identically distributed and
uncorrelated for sufficiently large temporal translations. Note the
constraints of the noise components in our null hypothesis are stronger than
that of Theiler's algorithm, which requires the noise distribution only
periodically depends on the phase of the signal. However, under our
hypothesis, we can produce the surrogates in a simple way through the
algorithm to be described below. In addition, a large scope of noise
processes often encountered in practical situations, including (but not
limited to) linear colored additive observational noise described by the $%
ARMA(p,q)$ model \cite{Box time}, match the above constraints.

The remainder of this communication is organized as follows. In Sec. II we
will introduce the new algorithm to generate pseudoperiodic surrogates,
while in Sec. III we will apply this algorithm to simulation data sets from
the R\"{o}ssler system, which demonstrates the ability of the surrogate test
based on this algorithm to distinguish chaotic orbits from pseudoperiodic
ones. As one of the applications, we will use this surrogate technique to
investigate whether a human electrocardiogram (ECG) record is possibly
presentative of a chaotic dynamical system. Finally, in Sec. IV, we will
have a summary of the whole communication.

\section{A New Algorithm to Generate Pseudoperiodic Surrogates}

%
%
\begin{figure}
\centering
\includegraphics[width=3.5in]{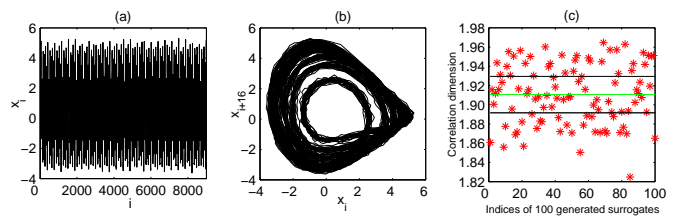}

\parbox{3.5in}{
\caption{\label{rosslerP5perObvDim4} (a) Pseudoperiodic time series contaminated
by observational noise;
(b) State space $x_{i+n}$ vs. $x_{i}$ of the pseudoperiodic time series from the
R\"{o}ssler system with $n=16$;
(c) Surrogate test for the pseudoperiodic time series based on our algorithm.
The abscissa is the indices
of 100 surrogates and the ordinate is the corresponding correlation dimensions.
The middle line is the mean correlation dimension of the original time series
calculated $100$ times using the GKA, the upper and lower lines denote the
correlation dimensions twice the standard deviation away from the mean value
and the asterisks indicate the correlation dimensions of $100$ surrogates. }
}

\end{figure}%

Let $\left\{ x_{i}\right\} _{i=1}^{N}$ be a data set with $N$ observations
(the form $\left\{ x_{i}\right\} $ is adopted instead for convenience when
causing no confusion), where $x_{i}$ means the observation measured at time $%
t_{i}=i\cdot \Delta t_{s}$ with $\Delta t_{s}$ denoting the sampling time.
We assume $\left\{ x_{i}\right\} _{i=1}^{N}$ is stationary and can be
decomposed into the deterministic components and the noise components, which
are approximately independent of each other. Similar to the surrogate test
idea of time shifting to desynchronize two data sets \cite{quiroga
performance}, we assume the noise components (approximately) follow an
identical distribution and are uncorrelated for sufficiently large temporal
translations (or time shifts). According to the null hypothesis we proposed
in the previous section, if the deterministic components are periodic, then
we can write a data point $x_{i}$ as $x_{i}=p_{i}+n_{i}$, where $p_{i}$ and $%
n_{i}$ denote the periodic component and the noise component respectively.
In many cases, we can set $E(p_{i})=E(n_{i})=0$ where $E$ is the expectation
operator. Since $\left\{ p_{i}\right\} $ are roughly independent of $\left\{
n_{i}\right\} $, we have the autocovariance $%
var(x_{i})=var(p_{i})+var(n_{i}) $. Let 
\begin{equation}
y_{i}^{\tau }=\alpha x_{i}+\beta x_{i+\tau }=(\alpha p_{i}+\beta p_{i+\tau
})+(\alpha n_{i}+\beta n_{i+\tau })  \label{linear combination}
\end{equation}%
with $i=1,2,~...,N-\tau $, where coefficients $\alpha $ and $\beta $ satisfy 
$\alpha ^{2}+\beta ^{2}=1$ and parameter $\tau $ is the temporal translation
between subsets $\left\{ x_{i}\right\} _{i=1}^{N-\tau }$ and $\left\{
x_{i+\tau }\right\} _{i=1}^{N-\tau }$, then the autocovariance function $%
var(y_{i}^{\tau })=var(\alpha p_{i}+\beta p_{i+\tau })+var(\alpha
n_{i}+\beta n_{i+\tau })$. Now let us consider the noise components. If $%
\tau $ is sufficiently large, under our hypothesis, $n_{i}$ and $n_{i+\tau }$
are uncorrelated. We also note that $\left\{ n_{i}\right\} $ and $\left\{
n_{i+\tau }\right\} $ are drawn from (approximately) the same distribution,
we have $var(\alpha n_{i}+\beta n_{i+\tau })=var(n_{i})$. For the
deterministic component, if we require the translation $\tau $ to satisfy $%
cov(p_{i},p_{i+\tau })=0$, then $var(\alpha p_{i}+\beta p_{i+\tau
})=var(p_{i})$. Hence by choosing a suitable temporal translation, the noise
levels of $\left\{ y_{i}^{\tau }\right\} $, defined by $\left( var(\alpha
n_{i}+\beta n_{i+\tau })/var(y_{i}^{\tau })\right) ^{1/2}$, will be the same
as that of $\left\{ x_{i}\right\} _{i=1}^{N}$, i.e., $\left(
var(n_{i})/var(x_{i})\right) ^{1/2}$. The reason to preserve the noise level
is that, the presence of noise will affect the calculation of the
correlation dimension, hence we would like to let the surrogates and the
original time series (roughly) have the same noise level in order to make
the results more conceivable.

%
%
\begin{figure}
\centering
\includegraphics[width=3.5in]{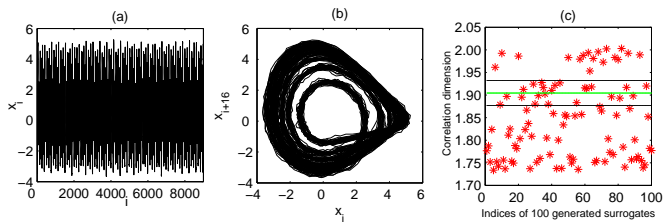}

\parbox{3.5in}{
\caption{\label{rosslerP5perObv015DDim4} (a) Pseudoperiodic time series
with both observational noise and dynamical noise;
(b) State space $x_{i+n}$ vs. $x_{i}$ of the pseudoperiodic time series from
the R\"{o}ssler system with $n=16$;
(c) Surrogate test for the pseudoperiodic time series based on our algorithm.
The meaning of the lines
and the asterisks is the same as that in panel $(c)$ of Fig. \ref{rosslerP5perObvDim4}. }
}

\end{figure}%

The above deduction leads to the central idea of our surrogate algorithm.
From Eq. (\ref{linear combination}), we note that if $\left\{ p_{i}\right\} $
is periodic, the nonconstant deterministic components $\left\{ \alpha
p_{i}+\beta p_{i+\tau }\right\} $ shall also be periodic. In addition, $%
\left\{ x_{i}\right\} _{i=1}^{N}$ and $\left\{ y_{i}^{\tau }\right\} $ shall
have the same noise level if a suitable translation $\tau $ is selected.
Therefore by randomizing the coefficient $\alpha $ or $\beta $, we can
generate many data sets $\left\{ y_{i}^{\tau }\right\} $ as the surrogates
of $\left\{ x_{i}\right\} _{i=1}^{N}$. Note that $\left\{ p_{i}\right\} $
and $\left\{ \alpha p_{i}+\beta p_{i+\tau }\right\} $ have the same
degree-of-freedom, if both of them are periodic, their correlation
dimensions \cite{Grassberger characterization} will theoretically be the
same. Now let us consider the noise components. Although the noise
components $\left\{ \alpha n_{i}+\beta n_{i+\tau }\right\} $ may have a
different distribution from that of $\left\{ n_{i}\right\} $, the noise
level is preserved after the transform in Eq. (\ref{linear combination}). As
Diks \cite{Diks estimating} has reported, the Gaussian kernel algorithm
(GKA) can reasonably estimate the correlation dimensions of noisy data sets
with different noise distributions. This implies that, under the same noise
level, the correlation dimensions of $\left\{ x_{i}\right\} _{i=1}^{N}$ and $%
\left\{ y_{i}^{\tau }\right\} $, calculated by the GKA, shall statistically
be the same if $\left\{ x_{i}\right\} _{i=1}^{N}$ and $\left\{ y_{i}^{\tau
}\right\} $ are both pseudoperiodic (and satisfy the constraints we
imposed). In contrast, if $\left\{ p_{i}\right\} $ is chaotic, its linear
combination, $\left\{ \alpha p_{i}+\beta p_{i+\tau }\right\} $, may have a
new dynamical structure with a different correlation dimension from that of $%
\left\{ p_{i}\right\} $, hence by adopting the correlation dimension as the
discriminating statistic we might detect this difference.

We shall also note that, for an unstable periodic orbit, even a small
dynamical noise might drive the resultant orbit far away from the original
position after a sufficiently long time, and the pseudoperiodicity might be
broken. In such situations, our algorithm might fail to work. Nevertheless,
we suggest to apply our algorithm as the first step in pseudoperiodicity
test, which is computationally fast and in principle deals well with a large
scope of observational noise (comparatively, the PPS algorithm will
sometimes fail for colored observational noise). If we can reject the null
hypothesis proposed previously, the time series in test is possibly chaotic
or pseudoperiodic perturbed by dynamical noise. Then we can adopt the PPS
algorithm for further tests, which works well even under a large amount of
dynamical noise. If \ the corresponding null hypothesis, i.e., the time
series is pseudoperiodic perturbed by dynamical noise, can be rejected
again, then we may claim the time series is very likely to be chaotic.

We now consider several computational issues in our algorithms. As described
in Eq. (\ref{linear combination}), to generate the surrogates $\left\{
y_{i}^{\tau }\right\} $, we select two subsets of $\left\{ x_{i}\right\}
_{i=1}^{N}$, $\left\{ x_{i}\right\} _{i=1}^{N-\tau }$ and $\left\{ x_{i+\tau
}\right\} _{i=1}^{N-\tau }$, multiply them by the coefficients $\alpha $ and 
$\beta $ respectively and then add them together. We shall emphasize that
choosing the temporal translation $\tau $ is a crucial issue for our
algorithm. From one aspect, we require the translation $\tau $ to satisfy
the condition $cov(p_{i},p_{i+\tau })=0$. The reason is that we want to keep
the noise level for the original time series and the surrogates. In
addition, we want the deterministic components $\left\{ \alpha p_{i}\right\} 
$ to be orthogonal to $\left\{ \beta p_{i+\tau }\right\} $ for arbitrary
coefficients $\alpha $ and $\beta $, otherwise the projection of $\left\{
\alpha p_{i}\right\} $ onto $\left\{ \beta p_{i+\tau }\right\} $ might
counteract $\left\{ \beta p_{i+\tau }\right\} $ under some situations, for
example, if $p_{i}\approx -p_{i+\tau }$ and $\alpha =\beta $, the
deterministic components $\left\{ \alpha p_{i}+\beta p_{i+\tau }\right\} $
will almost vanish while the noise components $\left\{ \alpha n_{i}+\beta
n_{i+\tau }\right\} $ remain. Hence the correlation dimensions calculated
are actually those of the noise components instead of the deterministic
components, which will certainly cause the false rejection of the null
hypothesis. From another aspect, we require $\tau $ to be sufficiently large
to guarantee the decorrelation between the noise components. However, we
expect $\left\{ x_{i}\right\} _{i=1}^{N-\tau }$ and $\left\{ x_{i+\tau
}\right\} _{i=1}^{N-\tau }$ shall have at least some overlaps to make use of
the information of the whole data set $\left\{ x_{i}\right\} _{i=1}^{N}$,
which means $\tau $ shall not exceed $N/2$. In addition, it is recommended
the length of a data set shall not be too short in order to appropriately
calculate its correlation dimension \cite{Jedynak failure}, which also
implies $\tau $ shall not be too large.

%
%
\begin{figure}
\centering
\includegraphics[width=3.5in]{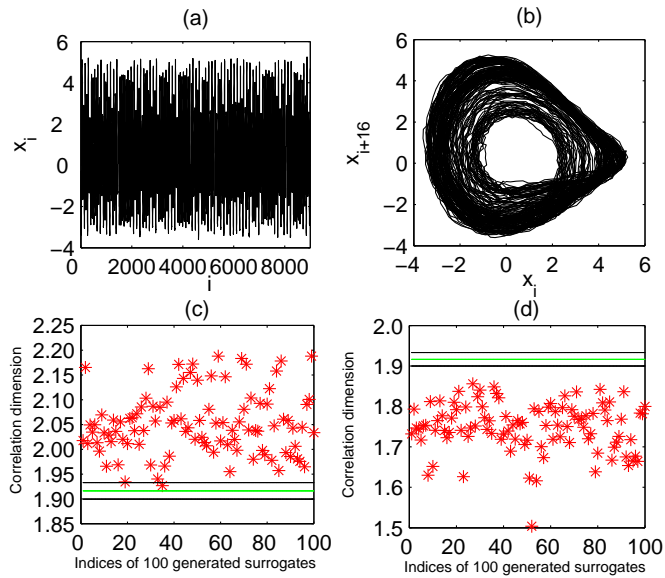}

\parbox{3.5in}{
\caption{\label{rossleCh5perObvDim5_pps} (a) Chaotic time series
contaminated by observational noise ;
(b) State space $x_{i+n}$ vs. $x_{i}$ of the chaotic time series from
the R\"{o}ssler system with $n=16$;
(c) Surrogate test for the chaotic time series based on our new algorithm.
The meaning of the lines
and the curve is the same as that in panel $(c)$ of Fig. \ref{rosslerP5perObvDim4};
(d) Surrogate test for the chaotic time series based on the PPS algorithm. The meaning
of the lines and the asterisks is the same as that in panel $(c)$ of Fig.
\ref{rosslerP5perObvDim4}.
}
}

\end{figure}%

From Eq. (\ref{linear combination}) we see that the surrogates are generated
from two segments $\{x_{i}\}_{i=1}^{N-\tau }$ and $\{x_{i+\tau
}\}_{i=1}^{N-\tau }$ of the original time series $\{x_{i}\}_{i=1}^{N}$. We
want segments $\{x_{i}\}_{i=1}^{N-\tau }$ and $\{x_{i+\tau }\}_{i=1}^{N-\tau
}$ to equivalently affect the generation of the surrogates, therefore we
would like to let $\max (\left\vert \alpha /\beta \right\vert )=\max
(\left\vert \beta /\alpha \right\vert )$, $\min (\left\vert \alpha /\beta
\right\vert )=\min (\left\vert \beta /\alpha \right\vert )$ and $\Pr
(\left\vert \alpha /\beta \right\vert \geqslant 1)\simeq \Pr (\left\vert
\beta /\alpha \right\vert \geqslant 1)$, where $\max (\cdot )$, $\min (\cdot
)$ and $\Pr (\cdot )$ denote the maximal function, the minimal function and
the probability function respectively. But note that the coefficient ratio $%
\alpha /\beta $ (or $\beta /\alpha $) shall not be too large or too small,
otherwise $\left\{ y_{i}^{\tau }\right\} $ will be very close to $\left\{
x_{i}\right\} _{i=1}^{N-\tau }$ or $\left\{ x_{i+\tau }\right\}
_{i=1}^{N-\tau }$, which will lead to approximately the same correlation
dimensions of $\left\{ x_{i}\right\} _{i=1}^{N}$ and $\left\{ y_{i}^{\tau
}\right\} $ regardless of the dynamical behavior of $\left\{ x_{i}\right\}
_{i=1}^{N}$, and thus decrease the discriminating power of the correlation
dimension. In our calculations we let $\alpha $ be uniformly drawn from the
interval $\left[ -0.8,-0.6\right] \cup \left[ 0.6,0.8\right] $ and $\beta =%
\sqrt{1-\alpha ^{2}}$, which satisfies our requirements and provides
moderate values for the ratio $\alpha /\beta $.

\section{Surrogate Test to Distinguish between Chaotic and Pseudoperiodic
Time Series}

In this section, through four examples from the R\"{o}ssler system, we
demonstrate the ability of surrogate test based on our algorithm to
discriminate chaotic orbits from pseudoperiodic ones. As an application, we
will also employ the surrogate technique to investigate whether a recorded
human electrocardiogram (ECG) data set is possibly chaotic.

\subsection{EXAMPLES}

The equations of the R\"{o}ssler system are given by 
\begin{equation}
\left\{ 
\begin{array}{l}
\dot{x}=-y-z, \\ 
\dot{y}=x+ay, \\ 
\dot{z}=b+z(x-c).%
\end{array}%
\right.  \label{rossler}
\end{equation}%
with the initial conditions $x(0)=y(0)=z(0)=0.1$. We choose parameters $b=2$%
, $c=4$ and the sampling time $\Delta t_{s}$ $=0.1$ time units. For each
example, the system is to be integrated $10,000$ times and the first $1,000$
data points are discarded to avoid including transient states.

In the first example, we set parameter $a=0.39095$. The R\"{o}ssler system
exhibits limit cycle behavior of period 6. To obtain pseudoperiodic time
series, we introduce $5\%$ observational noise into the periodic time
series. Although Gaussian white observational noise is the most common
choice in this situation, in order to demonstrate the ability of our
surrogate algorithm to deal with colored noise, we will instead adopt the
noise generated from the $AR(1)$ process \cite{Box time} $\xi _{i+1}=0.8\xi
_{i}+\epsilon _{i}$ with the variable $\epsilon $ following the normal
Gaussian distribution $N(0,1)$, which is the more difficult case due to the
correlation between noise components. However, one shall note that, Gaussian
white noise and other colored noises satisfying the constraints in our null
hypothesis, for example, those modelled by the $ARMA(p,q)$ processes, in
principle can be dealt with in the same way. For convenience of observation
and comparison, we plot the time series and the corresponding attractor in
two dimensional state space (or embedding space) in panels $\left( a\right) $
and $\left( b\right) $ of Fig. \ref{rosslerP5perObvDim4} respectively.

To produce surrogate data, first we shall choose a suitable temporal
translation. Since it is impractical to separate noise from signal
completely, in general it is difficult to estimate the correlation decay
time between noise components. Fortunately, to decorrelate noise components,
all temporal translations are equivalent as long as they are large enough.
In addition, in many real situations, it is often possible to observe the
background noise and thus estimate the decay time. In our example, we think
the $AR(1)$ noise to be uncorrelated when the temporal translation is larger
than $50$ (in units of the sampling time $\Delta t_{s}$). As another
requirement, temporal translation satisfying $cov(p_{i},p_{i+\tau })=0$ is
desired. In practice, of course, this requirement is generally impractical
due to the digitization and quantization in sampling process. Recall the
discussion in the previous section, by letting $E(p_{i})=0$ and $\alpha
^{2}+\beta ^{2}=1$, we have $var(\alpha p_{i}+\beta p_{i+\tau
})=var(p_{i})+2\alpha \beta \cdot cov(p_{i},p_{i+\tau })$. Function $%
cov(p_{i},p_{i+\tau })\neq 0$ means we do not preserve the noise level.
However, under the null hypothesis of pseudoperiodicity, there shall always
be some temporal translations to make $cov(p_{i},p_{i+\tau })\sim 0$, hence
the noise level will not deviate from the original one too much. Besides,
according to Eq. (\ref{linear combination}), we generate the surrogates by
uniformly drawing coefficient $\alpha $ from interval $\left[ -0.8,-0.6%
\right] \cup \left[ 0.6,0.8\right] $ ($\beta =\sqrt{1-\alpha ^{2}}$ is
always kept positive), the noise level of the surrogates will fluctuate
around that of the original one due to the alternative signs of product $%
\alpha \beta $. Therefore, $cov(p_{i},p_{i+\tau })\neq 0$ will only cause
some fluctuations when to calculate the correlation dimension because of the
fluctuations of noise level, however, generally such fluctuations will not
affect our conclusion if we can select a temporal translation $\tau $ to let 
$cov(p_{i},p_{i+\tau })\sim 0$. Since we have assumed the noise components
are roughly independent of the deterministic components, then $%
cov(x_{i},x_{i+\tau })=cov(p_{i},p_{i+\tau })$ for a large enough temporal
translation (to decorrelate noise components), therefore in all of the
examples, in order to let $cov(p_{i},p_{i+\tau })\sim 0$, we can
equivalently require $cov(x_{i},x_{i+\tau })\sim 0$. In the first example,
there are many temporal translations satisfying the two constraints
discussed above, i.e., $\tau >50$ and $cov(x_{i},x_{i+\tau })\sim 0$. To
pick a value from all these candidates, we first select an interval $\left[
100,150\right] $, then search the temporal translation which makes the
absolute value $\left\vert cov(x_{i},x_{i+\tau })\right\vert $ be the
minimum (most close to zero) among all translations $100\leqslant \tau
\leqslant 150$. One shall note that the choice of the interval $\left[
100,150\right] $ is arbitrary, except that we have to make sure that the
lower bound of the interval is larger than $50$, and there exists temporal
translations to let $cov(x_{i},x_{i+\tau })\sim 0$ within the interval.
After selecting the temporal translation, by randomizing the coefficient $%
\alpha $ we will generate $100$ surrogates according to Eq. (\ref{linear
combination}).

In order to calculate the correlation dimension, we adopt the time delay
embedding reconstruction \cite{Takens detecting} to recover the underlying
system from the scalar time series. Two parameters, i.e., embedding
dimension and time delay, shall be properly chosen to apply this technique.
Throughout this communication, we will use the false nearest neighbour
criterion \cite{kernel determining} to determine the global optimal
embedding dimension. Using the program in TISEAN package \cite{hegger
practical}, the embedding dimension $m$ of the original time series is
selected at $4$, which is the minimal value to make the fraction of false
nearest neighbours be zero. To choose a suitable time delay, we will use the
algorithm of redundancy and irrelevance tradeoff exponent (RITE) proposed in 
\cite{Luo geometric}. This algorithm selects the time delay by searching the
optimal tradeoff between redundancy (due to too small time delay) and
irrelevance (due to too large time delay). As demonstrated, the RITE
algorithm can select equivalently suitable time delays compared to the
average mutual information (AMI) criterion \cite{fraser and swinney
independent}, however, its implementation is much simpler and the
computational cost is fairly low. Therefore in case of large data sets,
adopting the RITE algorithm can facilitate our calculations. In the first
example we generate $100$ surrogates, and for each surrogate we keep the
embedding dimension $m=4$ and use the RITE algorithm to choose the suitable
time delay for time delay reconstruction.

%
%
\begin{figure}
\centering
\includegraphics[width=3.5in]{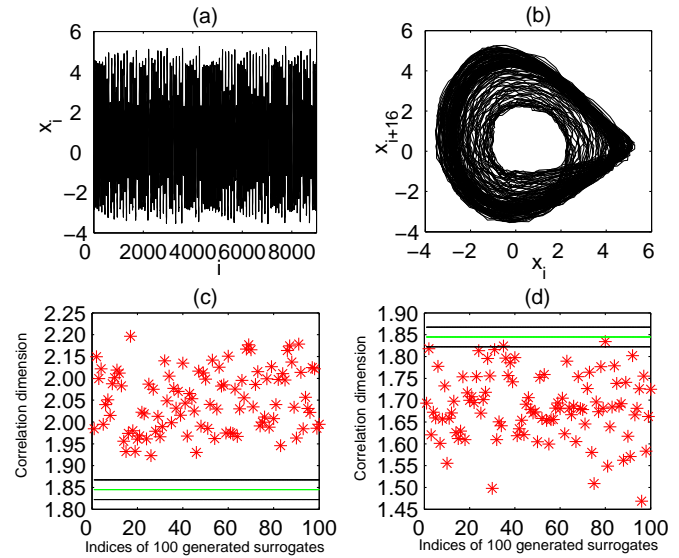}

\parbox{3.5in}{
\caption{\label{rossleCh5perObv015dDim4_pps} (a) Chaotic time series
with both dynamical and observational noises;
(b) State space $x_{i+n}$ vs. $x_{i}$ of the chaotic time series from
the R\"{o}ssler system with $n=16$;
(c) Surrogate test for the chaotic time series based on our new algorithm.
The meaning of the lines
and the asterisks is the same as that in panel $(c)$ of Fig. \ref{rosslerP5perObvDim4};
(d) Surrogate test for the chaotic time series based on the PPS algorithm. The meaning
of the lines and the asterisks is the same as that in panel $(c)$ of Fig.
\ref{rosslerP5perObvDim4}.
}
}

\end{figure}%

We will follow Diks's method \cite{Diks estimating} to calculate the
correlation dimension, which is more robust against noise by extending the
hard kernel function (or the Heaviside function) \cite{Grassberger
characterization} in calculation of correlation integral to the general
kernel functions. In his discussions, Diks adopted the Gaussian kernel
function, hence this method is called Gaussian kernel algorithm (GKA). Here
we will use the GKA implemented in \cite{Yu efficient} to calculate the
correlation dimensions, which further enhances the computational speed. Note
that to speed up the calculation, only 2000 data points are used as the
reference points for the GKA. There are some statistical fluctuations even
for the same data set when calculating its correlation dimension, therefore
for the original time series, we will calculate $100$ times to estimate the
mean correlation dimension and the standard deviation. As shown in panel $%
\left( c\right) $ of Fig. \ref{rosslerP5perObvDim4}, there are three lines
parallel to the abscissa. The middle line denote the estimation of the mean
correlation dimension of the original time series, while the upper and lower
lines indicate the positions twice the standard deviation away from the mean
value. For the surrogates, however, we will calculate their correlation
dimensions only once to save time. The results are illustrated as the
asterisks in panel $\left( c\right) $ of Fig. \ref{rosslerP5perObvDim4}.

After the calculation of the correlation dimensions, we need to inspect
whether the result is consistent with our null hypothesis. Here we use the
ranking criterion \cite{Theiler using} to determine whether the null
hypothesis shall be rejected or not. The idea of this criterion is that,
suppose the discriminating statistic of the original data set is $Q_{0}$,
and those of $N_{S}$ surrogates are $\left\{
Q_{1},Q_{2},~...,Q_{N_{S}}\right\} $. Rank the statistics $\left\{
Q_{0},Q_{1},~...,Q_{N_{S}}\right\} $ in the increasing order and denote the
rank of $Q_{0}$ by $r_{0}$, if the data set is consistent with the
hypothesis (i.e., no evidence to reject), $r_{0}$ can have an equal
possibility be any integer value between $1$ and $N_{S}+1$. However, if the
hypothesis is false, $Q_{0}$ might deviate from the surrogate distribution $%
\left\{ Q_{1},Q_{2},~...,Q_{N_{S}}\right\} $, i.e, $Q_{0}$ will be the
smallest or largest amongst $\left\{ Q_{0},Q_{1},~...,Q_{N_{S}}\right\} $,
hence we can reject the null hypothesis if $r_{0}=1$ or $N_{S}+1$, the
probability of a false rejection is $1/\left( N_{S}+1\right) $ for one-sided
tests and $2/\left( N_{S}+1\right) $ for two-sided tests.

For the first example, from panel $\left( c\right) $ of Fig. \ref%
{rosslerP5perObvDim4} we can see that, the mean correlation dimension of the
original time series falls within the dimension distribution of the
surrogates, therefore we cannot reject the null hypothesis as we expect,
which means the original time series is possibly pseudoperiodic \cite{note
interpretation}.

Now let us examine the other examples. In the second example, we still set
parameter $a=0.39095$ in Eq. (\ref{rossler}). However, to obtain the
pseudoperiodic time series, we first generate a data set by adding Gaussian
white noise with the standard deviation of $0.15\%$ to the $x$ component at
each integration step, which simulates the system perturbed by additive
dynamical noise, and then introduce $5\%$ observational $AR(1)$ noise into
the previously obtained data set. The global optimal embedding dimension is
chosen at $m=4$. Note in all of the four examples, we will generate $100$
surrogates, and parameter choices for surrogate generation will be the same,
i.e., we let the temporal translation be selected from $\left[ 100,150\right]
$ and coefficient $\alpha $ be uniformly drawn from $\left[ -0.8,-0.6\right]
\cup \left[ 0.6,0.8\right] $ ($\beta =\sqrt{1-\alpha ^{2}}$). For the second
example, the correlation dimensions of the original time series and the
surrogates are shown in panel $\left( c\right) $ of Fig. \ref%
{rosslerP5perObv015DDim4}. Under the ranking criterion, once again we cannot
reject our null hypothesis. Therefore the time series is possibly
pseudoperiodic, which is consistent with our knowledge.

In the third example, we change parameter $a$ of Eq. (\ref{rossler}) to be $%
0.395$. The R\"{o}ssler system exhibits chaotic behavior. We integrate Eq. (%
\ref{rossler}) to obtain a time series and then introduce $5\%$
observational $AR(1)$ noise. The optimal embedding dimension $m$ is selected
at $m=5$. From panel $\left( c\right) $ of Fig. \ref{rossleCh5perObvDim5_pps}%
, we find that the mean correlation dimension of the original time series
deviates from the distribution of the surrogate dimensions. Using the
ranking criterion, we can reject our null hypothesis. In order to exclude
the possibility that the time series is generated from an unstable period
orbit perturbed by dynamical noise, we also apply the PPS\ algorithm for
further test. From the PPS algorithm we generate $100$ surrogates, and then
use the GKA to calculate their correlation dimensions. The results are shown
in panel $\left( d\right) $ of Fig. \ref{rossleCh5perObvDim5_pps}, as we can
see, the mean correlation dimension of the original time series also falls
outside the distribution of the surrogate dimensions, therefore we can
reject the null hypothesis again. After the two surrogate tests for
pseudoperiodicity, we can claim the time series is chaotic with a confidence
level up to $96\%$ ($98\%\times 98\%$) for two-sided test.

%
%
\begin{figure}
\centering
\includegraphics[width=3.5in]{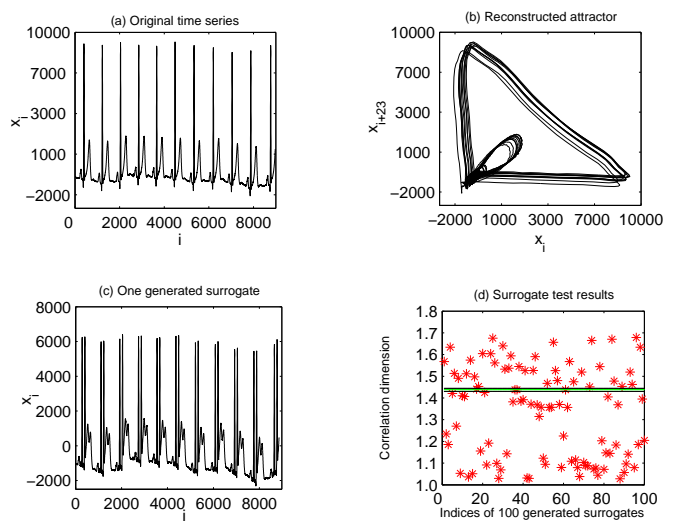}

\parbox{3.5in}{
\caption{\label{ecgtestEm5} (a) Time series of a human electrocardiogram
(ECG) record;
(b) State space $x_{i+n}$ vs. $x_{i}$ of the ECG record with $n=23$;
(c) A surrogate data generated from our algorithm with coefficient $\alpha=\beta=1/\sqrt{2}$ (cf. Eq. (\ref{linear combination}));
(d) Surrogate test for the ECG record based on our algorithm.
The meaning of the lines
and the asterisks is the same as that in panel $(c)$ of Fig. \ref{rosslerP5perObvDim4}. }
}

\end{figure}%

The final example to be demonstrated is a chaotic time series also from the R%
\"{o}ssler system. To generate the time series, we keep parameter $a=0.395$.
Similar to the way in the second example, we add Gaussian white noise with
the standard deviation of $0.15\%$ to the $x$ component at each integration
step as the dynamical noise, and then introduce $5\%$ observational $AR(1)$
noise into the previously obtained data set. The global optimal embedding
dimension is found to be $m=4$. The results of surrogate tests based on the
new algorithm and the PPS algorithm are shown in panel $(c)$ and $(d)$ of
Fig. \ref{rossleCh5perObv015dDim4_pps} respectively, from which we can see
that, surrogate tests based on both algorithms can detect the chaos in the
time series. Again we can claim the time series is chaotic with a confidence
level up to $96\%$ for two-sided test.

We have also investigated examples under different observational noise
levels (but keep the same dynamical noise if they have). For example, if we
reduce the $AR(1)$ observational noise levels to $3\%$ in the above four
examples, we can obtain the same results as we have reported. If we increase
the observational noise levels to $10\%$, for the pseudoperiodic time series
we can still obtain the expected results, i.e., we cannot reject our null
hypothesis. However, for the chaotic time series, we will falsely accept our
null hypothesis due to the correlation dimension of the original time series
marginally falling within the dimension distribution of the surrogates. The
reason of false acceptance might be that, under large noise levels, the
correlation dimension is not sensitive enough to detect the structure
changes of the chaotic time series. For such cases, we will have to look for
more powerful discriminating statistics \cite{schreiber dircrimination}.

\subsection{AN APPLICATION}

As an example of application, we employ the surrogate test based on our
algorithm to investigate whether a human electrocardiogram (ECG) record
(with $8975$ data points) is likely to be chaotic. The ECG record was
obtained by measuring from a resting healthy subject ($22$ years old) in a
shielded room at the sampling rate of $1$ KHz. The ECG record indicated in
panel $(a)$ of Fig. \ref{ecgtestEm5} looks very regular and even possibly
periodic, but we need a quantitative approach to verify the periodicity.
Here we choose the surrogate test technique. Using the false nearest
neighbour criterion, the global optimal embedding dimension is chosen at $%
m=5 $. The background noise is mainly from the measurement instruments,
usually it is a blend of white and correlated noise. By observing the linear
second order correlation function of the ECG data, we let the temporal
translation be within the interval $\left[ 100,150\right] $ (large enough to
decorrelate the noise components), where there exists an integer temporal
translation to make the correlation function almost be zero. Then by
uniformly drawing values from $\left[ -0.8,-0.6\right] \cup \left[ 0.6,0.8%
\right] $ for coefficient $\alpha $ in Eq. (\ref{linear combination}) ($%
\beta =\sqrt{1-\alpha ^{2}}$ ), we generate $100$ surrogates. For
demonstration, we plot in panel $\left( c\right) $ one surrogate generated
from Eq. (\ref{linear combination}) with coefficient $\alpha =\beta =1/\sqrt{%
2}$. We can see that the surrogate is different from the original ECG data
in that there appear more spikes in the surrogate. However, as we can also
find, the surrogate indicates the similar regularity to that in the original
data, which implies that the surrogate preserves the potential periodicity
in the original data as we expect (although in a different pattern). With
regards to the surrogate test, our calculation of the correlation dimensions
is also based on the GKA. The results are indicated in panel $\left(
d\right) $ of Fig. \ref{ecgtestEm5}, from which we can see that the mean
correlation dimension of the ECG data falls within the distribution of the
correlation dimensions of the surrogates, therefore we cannot reject our
null hypothesis. Hence the ECG record is possibly periodic. Moreover, there
is no evidence of chaos.

\section{Conclusion}

To summarize, by imposing a few constraints on the noise process, we devise
a simple but effective way to produce surrogates for pseudoperiodic orbits.
The main idea of this algorithm is that a linear combination of any two
segments of the same periodic orbit will generate another periodic orbit. By
properly choosing the temporal translation between the two segments, under
the same noise level we can obtain statistically the same correlation
dimensions of the pseudoperiodic orbit and its surrogates. Choosing the
temporal translation is a crucial issue for our algorithm, which in
principle shall guarantee the decorrelation between the noise components and
preserve the noise level. Another important issue is to select a proper
discriminating statistic which helps determine whether to reject the null
hypothesis. The correlation dimension is a suitable discriminating statistic
in this case.\ It is possible there are other suitable discriminating
statistics, we will leave the problem of finding such statistics for future
study.

The surrogate test technique based on our algorithm can be utilized to
distinguish between chaotic and pseudoperiodic time series. Initially, the
PPS algorithm was proposed to generate surrogates for a pseudoperiodic orbit
driven by dynamical noise, but sometimes surrogate tests based on this
algorithm will falsely reject the null hypothesis if the time series is also
contaminated by colored observational noise. As a complement to the PPS
algorithm, our algorithm deals well with observational noise, but it might
fail for large dynamical noise. However, due to the convenience in
computation, we suggest to adopt surrogate test based on our algorithm as
the first step for pseudoperiodicity detection. If we can reject the null
hypothesis of our algorithm, then we shall use the PPS algorithm for further
tests. If we can reject the null hypotheses of both the algorithms, then the
time series under investigation is very likely to be chaotic. In this
communication, the concrete procedures of surrogate test for
pseudoperiodicity are demonstrated through four simulation examples. As an
application in practice, we also employ the surrogate technique based on our
algorithm to investigate whether a human ECG record is possible to be
chaotic.

This research is supported by a Hong Kong University Grants Council
Competitive Earmarked Research Grant (CERG) number PolyU 5235/03E.


\begin{thebibliography}{10}
\bibitem[5]{note nonlinearity} Here we would like to elucidate that, the
irregularity of a time series is usually brought by stochasticity or
nonlinearity (often chaos), therefore in the example of nonlinearity
detection, if we can exclude (most of) the probability that the time series
is generated by a stochastic process, then it is very likely that the time
series is generated from a nonlinear deterministic system.
\bibitem[23]{note interpretation} We cannot claim the time series is
pseudoperiodic definitely, because if we cannot reject a null hypothesis,
there could be many interpretations other than that the data in test is
consistent with our null hypothsis. For more details, see, for example, Ref. 
\cite{Galka topics}.
\end{thebibliography}
\end{document}